\documentclass{iau} 
\usepackage{graphicx}
\usepackage{natbib}
\newcommand{\msun}{\mbox{$\,{\rm M}_\odot$}}

\title[Outer-Halo Globular Clusters] 
{Dynamical Evolution of Outer-Halo Globular Clusters}

\author[K\"upper et al.]   
{Andreas H.W. K\"upper$^{1}$
Akram H. Zonoozi$^2$, Hosein Haghi$^2$, Nora~L\"utzgendorf$^{\,3}$, Steffen Mieske$^4$, Matthias Frank$^5$, Holger~Baumgardt$^6$, and Pavel Kroupa$^7$
}

\affiliation{
$^1$Department of Astronomy, Columbia University, New York, NY 10027, USA \\email: {\tt akuepper@astro.columbia.edu}
\\[\affilskip]
$^2$Department of Physics, Institute for Advanced Studies in Basic Sciences, Zanjan, Iran\\[\affilskip]
$^{3}$ESA, Space Science Department, Keplerlaan 1, NL-2200 AG Noordwijk, The Netherlands\\[\affilskip]
$^4$European Southern Observatory, Alonso de Cordova 3107, Vitacura, Santiago, Chile\\[\affilskip]
$^5$Landessternwarte, Universit\"at Heidelberg, K\"onigsstuhl 12, 69117 Heidelberg, Germany\\[\affilskip]
$^6$University of Queensland, School of Mathematics and Physics, Brisbane, QLD 4072, Australia\\[\affilskip]
$^7$HISKP, Universit\"at Bonn, Nussallee 14-16, 53115 Bonn, Germany
}

\pubyear{2015}
\volume{316}  
\jname{Title of your IAU Symposium}
\editors{A.C. Editor, B.D. Editor \& C.E. Editor, eds.}
\begin{document}

\maketitle

\begin{abstract}
Outer-halo globular clusters show large half-light radii and flat stellar mass functions, depleted in low-mass stars. Using $N$-body simulations of globular clusters on eccentric orbits within a Milky Way-like potential, we show how a cluster's half-mass radius and its mass function develop over time. The slope of the central mass function flattens proportionally to the amount of mass a cluster has lost, and the half-mass radius grows to a size proportional to the average strength of the tidal field. The main driver of these processes is mass segregation of dark remnants. We conclude that the extended, depleted clusters observed in the Milky Way must have had small half-mass radii in the past, and that they expanded due to the weak tidal field they spend most of their lifetime in. Moreover, their mass functions must have been steeper in the past but flattened significantly as a cause of mass segregation and tidal mass loss. 

\keywords{methods: n-body simulations, globular clusters: general, Galaxy: halo}
\end{abstract}

\firstsection 

\section{Why are outer-halo globular clusters extended?}

\begin{figure}[t]
\begin{center}
 \includegraphics[width=2.2in]{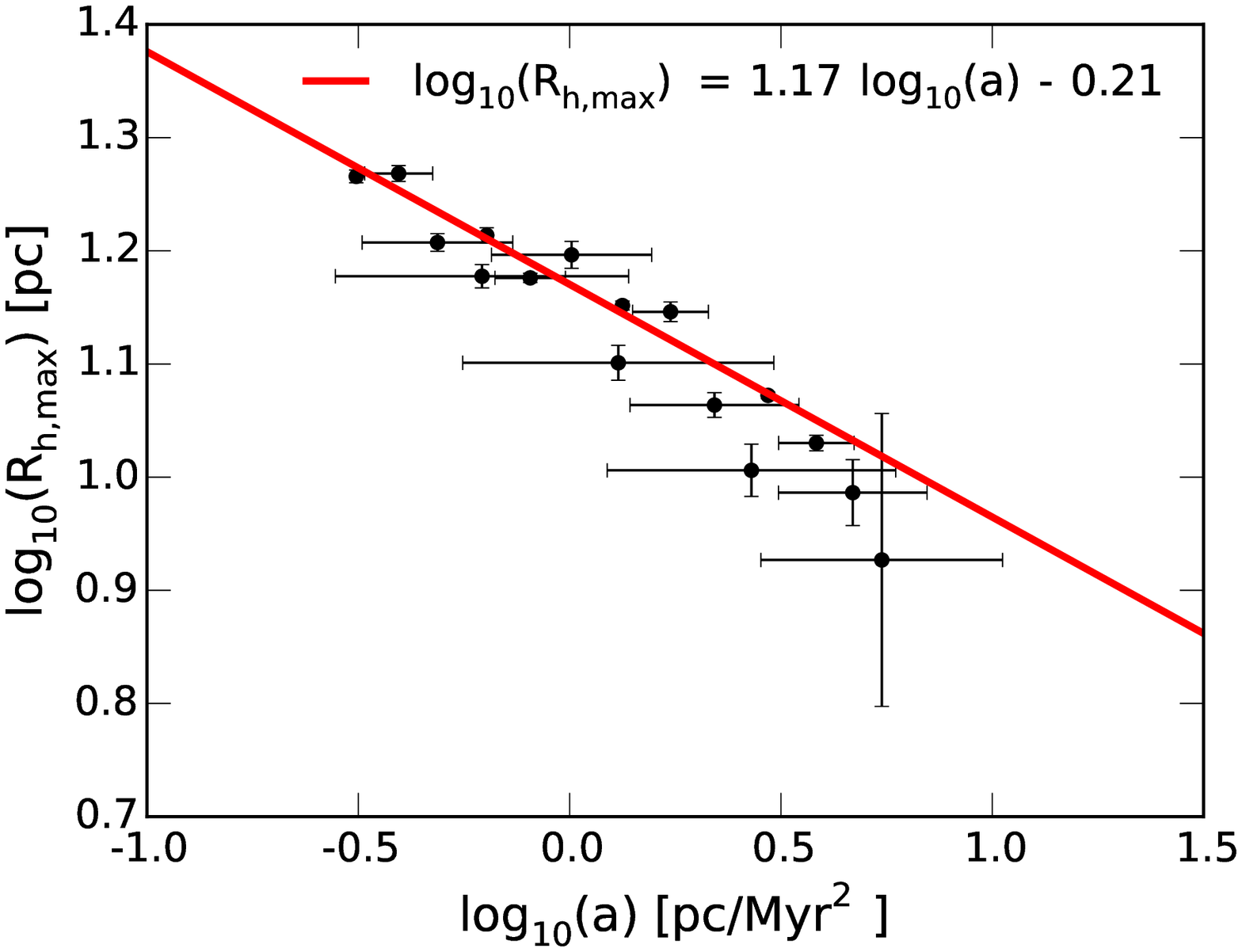} 
 \includegraphics[width=2.5in]{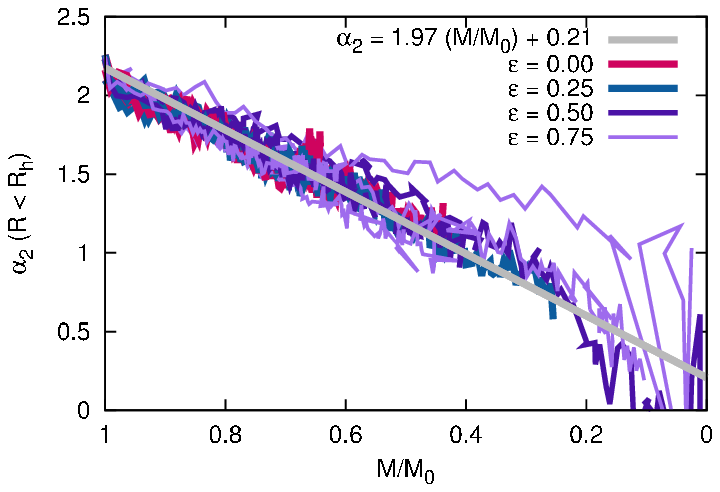} 
 \caption{\textit{Left:} the maximum half-mass radius, $R_{h,max}$, a cluster can reach throughout its evolution correlates with the time-averaged acceleration, $a,$ the cluster experiences on its orbit. \textit{Right:} the mass function slope of high-mass stars, $\alpha_2$, within the half-mass radius depends on the mass the cluster has lost independent of its orbit.}
   \label{fig1}
\end{center}
\end{figure}

Detailed photometric and spectroscopic investigations of outer-halo globular clusters (GCs) such as Palomar\,4 and Palomar\,14 have shown that these clusters are very extended, showing large half-light radii of $\gg10$\,pc, but at the same time show clear signatures of mass segregation within their centers (e.g., \citealt{Jordi09, Frank12, Frank14}). This is intuitively unexpected, since their present-day two-body relaxation times are larger than a Hubble time. One way of solving this apparent discrepancy is if these GCs were born compact, and the observed mass segregation happened in the early evolution of the clusters.

For the purpose of this investigation, we carried out 16 $N$-body simulations of the same globular cluster model in a fixed Milky Way-like galaxy potential using the collisional code \textsc{Nbody6} \citep{Aarseth03}. The GCs consisted initially of 65536 stars, following a \citet{Kroupa01} spectrum of stellar masses, and giving the clusters a mass of $M_0 = 26400\msun$. The GCs started off compact with an initial half-mass radius of 3\,pc. The models were placed on different orbits with a range of apogalactic distances, $r_{apo}=\{16, 32, 64, 128\}$\,kpc, and orbital eccentricities, $\epsilon = \{0, 0.25, 0.5, 0.75\}$. 

As has been observed in previous studies (e.g.,~\citealt{Baumgardt03}), the mass of the clusters, $M$, decreases faster with time for GCs orbiting in a stronger tidal field. Moreover, the initially compact clusters expand into their Roche lobes due to stellar mass loss and segregation of dark remnants into the cluster core. As a consequence of that, GCs in weaker tidal fields show larger half-mass radii, $R_h$, throughout their evolution (see, e.g., \citealt{Madrid12, Haghi14}). This expansion process can be reversed when the GCs have lost a significant amount of their initial mass. Hence, strongly dissolving clusters shrink after reaching a maximum expansion, $R_{h,max}$. In the left panel of Fig.~\ref{fig1}, we show the time-averaged tidal acceleration, $a$, and its variance of each GC plotted against the maximum half-mass radius. There is a clear correlation between the two quantities. Hence, the half-mass radii of outer-halo clusters reflect the tidal field they live in. Driver of the expansion are mainly dark remnants and high-mass stars segregating into the cluster core, and as such the degree of expansion depends on the initial composition of the cluster \citep{Haghi14}. As a consequence of this segregation, the mass-function slope, $\alpha_2$, of stars with $m>0.5\msun$ within the half-mass radius decreases proportionally to mass loss (see, e.g., \citealt{Vesperini97}). This flattening of mass function is independent of the orbit, as shown in the right panel of Fig.~\ref{fig1}. Deviations from this linear trend are only seen for clusters that undergo quick dissolution, and for $M/M_0<0.2$. We conclude that the flattened mass functions observed in the inner parts of extended outer-halo globular clusters are characteristic signatures of their expansion histories (cf., \citealt{Zonoozi11, Zonoozi14}).

\end{document}